\documentclass[11pt,a4paper]{article}
\usepackage{jcappub}

\usepackage{amsmath, amsfonts, amssymb, mathrsfs}
\usepackage{amsthm}
\usepackage{tensor}
\usepackage{multirow}
\usepackage{bbm} 
\usepackage[bb=boondox]{mathalfa}
\usepackage[geometry]{ifsym}

\usepackage{cmbright}

\makeatletter
\DeclareRobustCommand*{\bfseries}{%
  \not@math@alphabet\bfseries\mathbf
  \fontseries\bfdefault\selectfont
  \boldmath
}
\makeatother

\renewcommand{\nabla}{\!\!\mathrel{\raisebox{.15em}{%
           \reflectbox{\rotatebox[origin=c]{180}{$\triangle$}}}}\!\!} 

\newcommand{\La}{\mathcal{L}}
\newcommand{\Lab}{\mathcal{L}_{B}}
\newcommand{\Dslash}{{D\kern-0.63em{/}}}     
\newcommand{\scalf}{a}  
\newcommand{\J}{J_B} 
\newcommand{\Ps}{\mathit{\Psi}}
\newcommand{\Psb}{\bar{\mathit{\Psi}}}
\newcommand{\B}{B}
\newcommand{\gc}{\alpha}

\newcommand{\X}{\Omega}

\newcommand{\Y}[1]{\partial_{#1}\ln\X}
\newcommand{\N}{\left(\frac{\X - 1}{\X}\right)}
 
\newcommand{\D}{\mathit{\Delta}_{B}{}} 
 
\newcommand{\T}{\mathit{\Theta}}

\newcommand{\diracm}{\mathit{\Gamma}}
\newcommand{\coup}{\lambda}

\newcommand{\Nm}{\mu}
\newcommand{\Nn}{\nu}

\title{Gravitational baryogenesis without CPT violation}

\author[a]{Vicente Antunes,}
\author[b]{Ignacio Bediaga}
\author[a]{and Mario Novello}
 \affiliation[a]{Centro de Estudos Avan\c{c}ados de Cosmologia (CEAC), Rua Dr. Xavier Sigaud 150, Urca, CEP 22290-180,
 Rio de Janeiro, RJ, Brazil}
\affiliation[b]{Centro Brasileiro de Pesquisas F\'{i}sicas (CBPF), Rua Dr. Xavier Sigaud 150, Urca, CEP 22290-180, Rio de Janeiro, RJ, Brazil}
\emailAdd{antunes@cbpf.br}
\emailAdd{bediaga@cbpf.br}
\emailAdd{novello@cbpf.br}

\abstract{We show that in theories with nonminimal curvature-matter couplings up to first order in the curvature, baryon number ($B$) conservation and CP are automatically violated, provided that the curvature scalar and its gradient do not vanish. Together with the associated gravitationally induced particle creation in a cosmological framework, which is an essentially out-of-equilibrium process, this implies that all of Sakharov's conditions are satisfied in the simplest extension of General Relativity, without the need of CPT violation. Here, derivatives of the curvature can be disregarded, thus evading the potential problems associated with the Gravitational Baryogenesis (GB) model in a cosmological context. 
Thanks to the particle creation pressure and the trace anomaly, the curvature scalar is nonzero even in a radiation-dominated Universe. Moreover, the trace anomaly or the energy density of a massive scalar field can amplify the baryon asymmetry at very high temperatures, even when the particle creation rate is very slow. As a result, a baryon asymmetry 
compatible with observations and a huge CP violation can be generated in the early Universe.}


\arxivnumber{1909.03034}

\begin{document}

\maketitle


\section{Introduction \label{intro}}

\hspace{0.8cm} The excess of matter over antimatter in the Universe remains one of the greatest mysteries of modern cosmology, despite 50 years of continuous theoretical effort. 
The fact that no gamma-ray flux compatible with matter-antimatter annihilation in astrophysical systems have ever been detected strongly suggest that the observable patch of the Universe consists almost exclusively of matter. According to fundamental theories of physics, however, the Universe should be matter-antimatter symmetric to a very high degree \cite{GOLDHABER_1956}.
Precision measurements of the Cosmic Microwave Background (CMB) radiation \cite{PLANCK_2015} and predictions from primordial nucleosynthesis \cite{KOLB_TURNER, BBN_2015, PDG_NUCLEOSYN} provide a quantitative measure of this asymmetry, \textit{viz.}
\begin{equation}
\D \equiv \frac{n_{B} }{s} 
\simeq 10^{-10},
\label{delta}
\end{equation} 
where $n_{B} \equiv n_{b} - n_{\bar{b}}$ is the difference between the number densities of baryons and antibaryons, and $s$ is the entropy density.
This value is far above the expected relic abundance in a baryon-symmetric Universe from purely statistical fluctuations in baryon and antibaryon distributions ($\D \sim 10^{-39}$) \cite{ALPHER_HERMAN_1953, ALPHER_HERMAN_1958}.
 It has been suggested that the Universe may be partitioned in vast ($>$ 100 Gly) matter and antimatter domains \cite{COHEN_DERUJULA_GLASHOW_1998, DOLGOV_1997, CLINE_2006}, but no mechanism capable to explain the formation of such domains is known (see, however, \cite{DOLGOV_ET_AL_2015}). 
 
This impasse sparked the idea that the baryon asymmetry must have been generated dynamically in an early phase of the expanding Universe. In a seminal paper, Sakharov established the conditions for the generation of this asymmetry from symmetrical initial conditions, \textit{viz.} \cite{SAKHAROV_1966}:
\begin{enumerate}
\item[(i)] Baryon number ($\B$) nonconservation;
\item[(ii)] Charge conjugation (C) and combined charge conjugation and parity transformation (CP) violations;
\item[(iii)] Departure from thermal equilibrium. 
\end{enumerate}
The last condition is usually attributed either to the decay of superheavy particles or to cosmological first-order phase transitions.

In the context of the Standard Cosmological Model (SCM) and the Standard Model of Particle Physics (SM), there are three relevant scales for this problem. (1) Any baryon asymmetry produced above the Grand Unification (GUT) scale ($\sim 10^{16}$ GeV) would be washed away by cosmic inflation, thus ruling out a GUT baryogenesis. 
(2) Despite the fact that $B$ can be violated at the electroweak (EW) scale ($T\gtrsim 100$ GeV) \cite{ABJ1,ABJ2,THOOFT,KUZMIN_RUBAKHV_SHAPOSHNIKOV_1985, SHAPOSHNIKOV_1986, SHAPOSHNIKOV_1987}, 
a first order EW phase-transition is ruled out by the observed Higgs boson mass \cite{KAJANTIE_ET_AL_1996}.
Moreover, although C is maximally violated in the weak interaction, the CP-violation is not sufficient to account for the observed asymmetry \cite{CLINE_2006, RIOTTO_1998}. 
(3) Similarly, lattice Quantum Chromodynamics (QCD) simulations favour a sharp cross-over rather than a phase transition in full QCD (for a null chemical potential), while the CP violation expected in the most general QCD Lagrangian, 
  \textit{viz}.
\begin{equation}
\La_{QCD} \supset \theta\mbox{tr}\tensor{F}{_{\Nm\Nn}}\tensor{\tilde{F}}{^{\Nm\Nn}} + m\Psb e^{i\theta'\gamma^{5}}\Ps,
\end{equation}
was never observed (strong CP problem). Furthermore, no sources of $B$ violation are known in QCD. In face of these difficulties, many baryogenesis mechanisms based on SM extensions have been proposed in the last decades (see  \cite{ DOLGOV_ZELDOVICH_1981, DOLGOV_1993, DOLGOV_1997, CLINE_2006, RIOTTO_1998, RIOTTO_TRODDEN_1999} for reviews), but a solution to the problem is still lacking.
 
On the other hand, since the work of Sakharov there has been a growing suspicion that gravity might play an essential role in brayogenesis due to the cosmological nature of this process. This picture can be linked to a cosmological influence on the fundamental interactions, an idea which has now a long history \cite{DIRAC_1937_a,NOVELLO_ROTELLI_1972}. In particular, such an influence is expected in higher-order modifications of General Relativity (GR), defined by the Lagrangian
\begin{equation}
\La_{G} =  \frac{M_P^2}{2}R + \varepsilon_1 R^2 + \varepsilon_2 R_{\Nm\Nn}R^{\Nm\Nn} 
+ \cdots , \label{pre_lagrangian_1}
\end{equation}
where $M_P= (8\pi G)^{-1/2}\simeq 2.4\times 10^{18}$ GeV is the reduced Planck mass, 
$\varepsilon_i$ are dimensionless constants, and throughout the text we adopt the natural units $k_B = \hslash = c =1$. In such a theory, the matter sector Lagrangian should also be written with general couplings as an expansion in powers of the curvature 
\begin{equation}
\La_M^{eff.} = \La_M + \lambda_1 R\La_M + \lambda_2R^{\Nm\Nn} \frac{\delta \La_M}{\delta g^{\Nm\Nn}} + \cdots,
\label{pre_lagrangian_2}
\end{equation}
where $\La_M$ is the ``pure'' matter Lagrangian, and the coupling constants $\lambda_i$ have dimensions of $[\mbox{mass}]^{-2}$. In the last decades, there has been considerable interest in such extensions of GR in connection with nonsingular cosmological models \cite{NOVELLO_SALIM_1979} (see \cite{NOVELLO_BERGLIAFFA_2008} for a review) and the ``dark sector'' (see \textit{e.g.} \cite{BBHL_2007, BM_2012,BISABR_2012}). Even if these couplings are irrelevant in the gravitational field of the present-day Universe, they might play a crucial role in the early Universe. This is the framework behind the popular model dubbed ``Gravitational Baryogenesis'' (GB), which relies on the CP-violating interaction $\La_M^{eff.}\supset \coup J^{\Nm}\partial_{\Nm}R$, where $J^{\Nm}$ is a fermion current, and which leads to dynamical $B$ and CPT violations \cite{STEINHARDT_ET_AL_2004}. Extensions of this model were proposed in \cite{LAMBIASE_2006, ODINSTOV_2016}. Although CPT violation dispenses with the third of Sakharov's conditions, the consistency of the cosmic dynamics in this scenario has been questioned \cite{ARBUZOVA_DOLGOV_2017,ARBUZOVA_DOLGOV_2017b}. In contrast, consistent cosmic dynamics are well known in theories with nonminimal curvature-matter couplings up to first order in $R$ \cite{NOVELLO_SALIM_1979}.

 As first discussed by Zeldovich, higher-order corrections of GR of the form (\ref{pre_lagrangian_1}) are by necessity associated with a spontaneous gravitationally induced particle creation process, since the nonzero divergence of the energy-momentum tensor of matter is a necessary condition for such a process to occur \cite{ZELDOVICH_1970}. 
A phenomenological description of a gravitationally induced particle creation process in a cosmological context was first provided by Zeldovich himself, who noted that this could be understood as a sort of ``viscosity of the vacuum'' which gives rise to a negative particle creation pressure in the effective energy-momentum tensor \cite{ZELDOVICH_1970}. Alternatively, a description of this process was provided by Prigogine et al. within the framework of irreversible thermodynamics of open systems \cite{PRIGOGINE_GEHENIAU_1986,PRIGOGINE_ET_AL_1988,PRIGOGINE_ET_AL_1989}, and further elaborated in \cite{CALVAO_LIMA_WAGA_1992}. According to this description, any such gravitationally induced particle creation process in an expanding Universe amounts to an irreversible energy flow from the gravitational field to created particle-antiparticle pairs, which is an essentially out-of-equilibrium process. In this framework, however, an equal number of particles and antiparticles is produced. Recently, a connection between general relativistic particle creation and baryogenesis in the context of the GB model was proposed \cite{LIMA_SINGLETON_2016}.

Alternatively, Harko et al. have described, from a phenomenological point of view, the gravitationally induced particle creation process associated with general theories $\La = f(R,\La_M)$ with nonminimal curvature-matter couplings \cite{HARKO_2014,HARKO_2015}. As shown by these authors, this follows directly from the nonconservation of the energy-momentum tensor of matter in such theories,  which, in the perspective of the thermodynamics of open systems, leads to an irreversible flow of energy from the gravitational field (curvature) to created matter-antimatter pairs in a homogeneous and isotropic cosmological setting.

 In the present work, we will examine the role of nonminimal curvature-matter couplings in the baryogenesis problem. For that purpose, we will consider nonminimal couplings between baryons (quarks) and gravity up to first order in the curvature only. We show that the field equations imply a nonconservation of the baryon current given by $\,\nabla_{\Nm}\J^{\Nm} \simeq -\coup \J^{\Nm}\partial_{\Nm}R$. Moreover, as suggested by QCD, a CP-violating term $i\coup R\Psb\diracm^{5}\Ps$ is naturally expected in the nonminimal interaction Lagrangian. Since the particle creation in an expanding Universe induced by the curvature-matter couplings is an essentially out-of-equilibrium process, all of Sakharov's conditions are automatically satisfied, without the need of CPT violation, provided that the curvature scalar and its gradient do not vanish. In our approach the potential problems of the GB model associated with higher-order derivatives of the curvature are evaded. By virtue of the particle creation pressure and the $SU(N_c)$ ``trace anomaly'' $\tensor{T}{_{\Nm}^{\Nm}} \propto \beta(\gc)\mbox{tr}\tensor{F}{_{\Nm\Nn}}\tensor{F}{^{\Nm\Nn}} \neq 0$, the later related to the dynamical breaking of the conformal invariance at very high temperatures, $R$ is nonzero even in a radiation-dominated Universe. Since $\partial_{\Nm}R  = \delta_{\Nm}^{\ 0}\dot{R}$ and $\dot{R} < 0$ in an expanding homogeneous and isotropic Universe, a baryon asymmetry can be produced. Moreover, the trace anomaly or the energy density of a massive scalar field can amplify the baryon asymmetry at very high temperatures, even when the particle creation rate is very slow.

\section{Nonminimal curvature-matter coupling and Sakharov's conditions}

\hspace{0.8cm} We consider the action
\begin{equation}
S = \int \La \sqrt{-g}\, d^4 x,
\label{action}
\end{equation}
defined by the simplest extension of the GR Lagrangian which describes fermions nonminimally coupled to gravity 
\begin{align}
\La = \frac{M_P^2}{2}R + \Lab + \lambda R\Lab + im\coup R\Psb\diracm^{5}\Ps + \La_{A,\phi,\cdots},
\label{lagrangian_2}
\end{align}
where $\Lab$ is the Dirac Lagrangian
\begin{equation}
\Lab \equiv \frac{i}{2}\left( \Psb \diracm^{\Nm}D_{\Nm} \Ps - D_{\Nm}\Psb\diracm^{\Nm}\Ps \right) - m\Psb\Ps,
\end{equation}
 and $\La_{A,\phi,\cdots}$ denotes the Lagrangian of the other forms of ``matter'' (including $SU(N_c)$ gauge fields), which, for simplicity, we assume to be minimally coupled to gravity. In the expression above, the spinor field $\Ps$ represents fermions, $m$ is the mass of the fermion, the conjugate spinor is defined as $\bar{\Ps} = \Ps^{\dagger}\gamma^{0}$, and summation over all fermion flavours was omitted. We assume that the fermions described here correspond to baryons (more specifically, quarks).
The flat space-times gamma matrices $\gamma^{a}$ and the tetrad fields $\tensor{e}{^{a}_{\Nm}}$ allows one to define the generalized Dirac matrices in curved space-time
\begin{equation}
\diracm^{\Nm} \equiv \gamma^{a}\tensor{e}{_{a}^{\Nm}},
\end{equation}
which satisfy the Clifford algebra $\{\diracm^{\Nm},\diracm^{\Nn}\} = \diracm^{\Nm}\diracm^{\Nn} + \diracm^{\Nn}\diracm^{\Nm} = 2g^{\Nm\Nn}I$, where $g_{\Nm\Nn} = \tensor{e}{^{a}_{\Nm}}\eta_{ab}\tensor{e}{^{b}_{\Nn}}$ is the space-time metric, $\eta_{ab}=\mbox{diag}(+1, -1, -1, -1)$ is the Minkowski metric, and $g\equiv \mbox{det}(g_{\Nm\Nn})$ is the determinant of the metric. The covariant derivative for a spinor field is defined in terms of the spin connection by 
\begin{equation}
D_{\Nm}\Ps \equiv \partial_{\Nm}\Ps + \frac{1}{8}(\tensor{e}{_{a}_{\Nn}}\nabla_{\Nm}\tensor{e}{_{b}^{\Nn}})[\gamma^a,\gamma^b ]\Ps,
\end{equation}
where we have disregarded, for simplicity, the interaction between baryons and other forms of matter. We also have defined $\diracm^5 \equiv i\gamma^{0}\gamma^{1}\gamma^{2}\gamma^{3} = \gamma^5$, such that $\{\diracm^{\Nm},\diracm^{5}\} = 0$.

Variation of the action (\ref{action}) with respect of the spinor fields produces the field equation for baryonic matter
\begin{equation}
i\diracm^{\Nm}D_{\Nm}\Ps - m\Ps = -\frac{i}{2}(\partial_{\Nm}\ln\X)\diracm^{\Nm}\Ps - im\left(\frac{\X - 1}{\X}\right)\diracm^{5}\Ps , 
\label{eq_dirac_1}
\end{equation}
where we have defined
\begin{equation}
\X(R) \equiv 1 + \coup R . 
\label{omega_factor0}
\end{equation}
We define the baryon current in curved space-time as  
\begin{equation}
\J^{\Nm} \equiv \Psb\diracm^{\Nm}\Ps.
\end{equation}
Taking the divergence of the baryon current, and combining the result with equation (\ref{eq_dirac_1}) and its conjugate, we obtain
\begin{equation}
\nabla_{\Nm}\J^{\Nm} = - (\Y{\Nm})\J^{\Nm}.
\label{current_cons_viol}
\end{equation}
Equation (\ref{current_cons_viol}) states the nonconservation of the baryon current in theories with nonminimal curvature-matter couplings, provided that the gravitational field is nonstationary. A comment is in order here. One can proceed canonically and define a conserved current $\tilde{J}_B^{\Nm} = \X\J^{\Nm}$. Such a definition, however, lacks a clear physical meaning, since it mixes fermions and the gravitational field. 
  
The next thing to note is that the Lagrangian (\ref{pre_lagrangian_2}) explicitly violates the CP-symmetry, since the imaginary mass term transforms under CP as 
\begin{equation}
\mbox{CP}: im\coup R\Psb\diracm^{5}\Ps\longrightarrow -im\coup R\Psb\diracm^{5}\Ps.
\end{equation}
Therefore, the second of Sakharov's conditions is automatically satisfied, provided that the curvature scalar does not vanish.

Finally, variation of the action (\ref{action}) with respect to the space-time metric yields the field equations of gravity
\begin{equation}
R_{\Nm\Nn} - \frac{1}{2}Rg_{\Nm\Nn} = -\frac{1}{M_P^2} \T_{\Nm\Nn}, 
\label{field_eq}
\end{equation}
with ``effective'' energy-momentum tensor
\begin{align}
\T_{\Nm\Nn} =\ & T_{\Nm\Nn} - XT_{\Nm\Nn} - M_P^{2}\left( \mathcal{P}_{\Nm\Nn}X -R\frac{\delta X}{\delta g^{\Nm\Nn}} \right) + \mathcal{O}^2(X),
\label{field_eq0}
\end{align}
where $T_{\Nm\Nn} = \sum_I T^I_{\Nm\Nn}$ is the total energy-momentum tensor of matter composed by the sum of the energy-momentum tensors
\begin{equation}
 T^{I}_{\Nm\Nn} \equiv \frac{2}{\sqrt{-g}}\frac{\delta(\sqrt{-g}\La_{I})}{\delta g^{\Nm\Nn}}
\end{equation}
 of each matter component $I=B,A,\phi,\cdots$, we have introduced $\mathcal{P}_{\Nm\Nn} \equiv g_{\Nm\Nn}\Box\, - \nabla_{\Nm}\!\nabla_{\Nn}$ for notational convenience, and we have defined
\begin{equation}
X \equiv 2\coup M_P^{-2}\left( \Lab + im\Psb\diracm^5\Ps \right).
\end{equation}
In particular, the energy-momentum tensor of baryonic matter has the explicit form
\begin{equation}
 T^{B}_{\Nm\Nn} = \frac{i}{2}\left[ \Psb \diracm_{(\Nm}D_{\Nn)} \Ps - D_{(\Nn}\Psb\diracm_{\Nm)}\Ps \right] - \Lab g_{\Nm\Nn},
 \label{energy_momentum_fermions}
\end{equation}
in which we have used the definition $2S_{(\Nm\Nn)} \equiv S_{\Nm\Nn} + S_{\Nn\Nm}$. Taking the divergence of equation (\ref{field_eq}), it follows that
\begin{equation}
\nabla^{\Nm}\T_{\Nm\Nn} = 0. \label{cons_theta}
\end{equation}
Note that the tensor $\T_{\Nm\Nn}$ cannot be separated in conserved contributions from each form of matter alone. Since the other forms of matter are minimally coupled to gravity, we can assume that $\nabla^{\Nm}T_{\Nm\Nn}^I = 0$ for all $I\neq B$. 
Equation (\ref{cons_theta}) implies the nonconservation of the energy-momentum tensor of baryonic matter, \textit{viz.}
\begin{equation}
\nabla^{\Nm}T^{B}_{\Nm\Nn}
 \neq 0.
\label{emt_noncons}
\end{equation}

The nonvanishing of the covariant divergence of the energy-momentum tensor has long been interpreted as suggesting a matter creation process in nonstationary gravitational fields \cite{ZELDOVICH_1970}. This process was studied by Harko et al., from a phenomenological point of view,  in the general framework of $f(R,\La_M)$ theories with nonminimal curvature-matter couplings \cite{HARKO_2014,HARKO_2015}. As we will discuss in the following section, these authors have shown that in the framework of irreversible thermodynamics of open systems, equation (\ref{emt_noncons}) implies an irreversible particle creation process, which amounts to the energy transfer from the gravitational field (curvature) to ``real'' matter. On the other hand, according to equation (\ref{current_cons_viol}), although the gravitationally induced particle creation is $B$-symmetric in the general relativistic process, the process associated with nonminimal curvature-matter couplings is essentially $B$-asymmetric.

\section{Cosmological particle creation and baryon asymmetry}

 \hspace{0.8cm} We now proceed to consider the consequences for baryogenesis of the particle creation induced by the curvature-matter (baryons) coupling in an expanding Universe. For that purpose, we assume a homogeneous and isotropic space-time described by the spatially flat Friedman-Lema\^{i}tre-Robertson-Walker (FLRW) metric
\begin{equation}
ds^2 = dt^2 - \scalf^2(t)\left( dx^2 + dy^2 + dz^2 \right),
\end{equation}
where $\scalf(t)$ is the scale factor. The corresponding tetrad fields are
\begin{equation}
\tensor{e}{^{0}_{\Nm}} = \tensor{\delta}{^{0}_{\Nm}} \ , \ \tensor{e}{^{i}_{\Nm}} = \scalf(t)\tensor{\delta}{^{i}_{\Nm}}.
\end{equation}
The generalized Dirac matrices, therefore, are given by $\diracm^0 = \gamma^0$ and $\diracm^{i} = \gamma^{i}/\scalf$. In this particular geometry, the energy-momentum tensor of a homogeneous fermion field $\Ps = \Ps(t)$ nonminimally coupled to gravity is compatible with spatial homogeneity and isotropy. This can be shown by calculating the components of the matter energy-momentum tensor (\ref{energy_momentum_fermions}), \textit{viz.}
\begin{equation}
T^{B}_{00} = m\Psb\Ps ,\label{emt_b_00}
\end{equation}
\begin{equation}
T^{B}_{ij} = im\N\Psb\gamma^{5}\Ps g_{ij},\label{emt_b_ij}
\end{equation}
\begin{equation}
T^{B}_{0i} = 0,
\end{equation}
in which we have used the field equation (\ref{eq_dirac_1}) and its conjugate, which in this geometry assumes the form
\begin{equation}
\dot{\Ps} + \frac{3}{2}H\Ps = - im\gamma^0\Ps -\frac{1}{2}(\Y{0})\Ps - m\N\gamma^0\gamma^{5}\Ps , 
\label{eq_dirac_1b}
\end{equation}
where $H=\dot{\scalf}/\scalf$ is the Hubble parameter, the overdot means time derivative, and 
\begin{equation}
\Y{0} = \frac{\coup \dot{R}}{1 + \coup R}.
\end{equation} 
Therefore, in this geometry the energy-momentum tensor of each matter component can be written in the form
\begin{equation}
T^I_{\Nm\Nn} =  (\rho_I + p_I )V_{\Nm}V_{\Nn} - p_I g_{\Nm\Nn},
\end{equation}
where $V^{\Nm} = \tensor{e}{_{0}^{\Nm}} = \delta^{\Nm}_{\ 0}$ is the co-moving vector field. For the baryons, in particular, we have $\rho_B = m\Psb\Ps$ and $p_B = -im\N\Psb\gamma^{5}\Ps$. The expansion (\ref{field_eq0}) can finally be used to show that $\T_{\Nm\Nn}$ is diagonal, \textit{i.e.} $\T_{0i} = 0$. Despite this fact, the non-vanishing of the spatial components of the corresponding homogeneous baryon current $\J^{\Nm} = \J^{\Nm}(t)$ may still break spatial isotropy. Although there are particular choices of the spinor field for which these components vanish, in general this is not the case. Such isotropy violations, however, are not detectable if the current does not couple to any observable vector field, as is the case in the Lagrangian (\ref{pre_lagrangian_2}). Note that this is not the case in the GB model, where the current couples to the vector field $\partial_{\Nm}R$.

  In terms of the homogeneous current $\J^{\Nm}= \J^{\Nm}(t)$, equation (\ref{current_cons_viol}) assumes the form
\begin{equation}
\dot{n}_{B} + 3Hn_{B} = - (\Y{0})n_{B},
\label{baryon_numb_production}
\end{equation}
where $n_{B}(t) = \J^0(t)$ is the baryon number density. Integration of expression (\ref{baryon_numb_production}) yields
\begin{equation}
n_{B0} = n_{B\ast}\left(\frac{\scalf_{\ast}}{\scalf_0} \right)^3 \left( \frac{1+ \coup R_{\ast}}{ 1 + \coup R_0} \right), 
\label{baryon_number_densit}
\end{equation}
where $n_{B\ast}$ and $n_{B0}$ are, respectively, the initial and final baryon number densities, and the same for the other quantities, $\scalf_{\ast} = \scalf(t_{\ast})$, \textit{etc}. 

According to Harko et al. \cite{HARKO_2014,HARKO_2015}, 
in the framework of the thermodynamics of open systems, the change in the internal energy of the system implied by equation (\ref{emt_noncons})
is entirely due to the change of the number of particles. In the cosmological context, this change is due to the irreversible transfer of energy from the gravitational field to created matter, where matter creation acts as a source of internal energy. 
The second law of thermodynamics 
implies the particle number balance equation \cite{PRIGOGINE_GEHENIAU_1986,PRIGOGINE_ET_AL_1988,PRIGOGINE_ET_AL_1989}
\begin{equation}
\dot{n} + 3Hn = \Gamma_cn,
\end{equation}
where $\Gamma_c$ 
is the particle creation rate induced by the nonminimal curvature-matter coupling. The exact form of the particle creation rate was obtained by Harko \cite{HARKO_2014} in the more general context of $f(R,\mathcal{L}_m)$ theories. This, however, will not be relevant to us here. It suffices to consider the general energy balance equation, which can be written as
\begin{equation}
\dot{\rho} + 3H(\rho + p + p_c) = 0,\label{cons}
\end{equation}
where $\rho = \sum_I \rho_I$ and $p = \sum_I p_i$ are, respectively, the energy-density and pressure of the cosmic fluid, and the matter creation pressure is given by \cite{PRIGOGINE_GEHENIAU_1986,PRIGOGINE_ET_AL_1988,PRIGOGINE_ET_AL_1989}
\begin{equation}
p_c 
= -\frac{\rho_B + p_B}{3H}\Gamma_c.
\end{equation}

According to equations (\ref{cons_theta}) and (\ref{cons}), an adiabatic gravitationally induced irreversible cosmological matter creation process in a homogeneous and isotropic space-time can be phenomenologically described by the effective energy-momentum tensor
\begin{equation}
\T_{\Nm\Nn} = (\rho + p + p_c)V_{\Nm}V_{\Nn} - (p + p_c)g_{\Nm\Nn},
\end{equation}
where $\T_{\Nm\Nn}\to T_{\Nm\Nn}$ as $p_c\to 0$. This description is formally equivalent to the particle creation process proposed by Zeldovich in the context of GR in terms of bulk viscosity of the cosmic fluid \cite{ZELDOVICH_1970}.
We assume that each matter component satisfies the barotropic equation of state $p_I = w_I\rho_I$. With these assumptions, the trace of the field equations of gravity (\ref{field_eq})
yields
\begin{equation}
R = \frac{1}{M_P^2} \left[ \sum_I (1 - 3w_I)\rho_I + (1 + w_B)\rho_B\frac{\Gamma_c}{H} \right] , \label{g_field_2b}
\end{equation}
where we have used $\tensor{\T}{_{\Nm}^{\Nm}} = \rho -3p - 3p_c$.

 Note that even for a slow particle creation rate, the asymmetry between created baryons and atibaryons can be strongly amplified by the contribution from other forms of matter, minimally coupled to gravity or not. 
Since the entropy production that results from the induced particle creation promotes a dilution of the baryon asymmetry $\D$ (in the adiabatic case, the entropy scales as $S\sim \scalf^{3})$ \cite{LIMA_GERMANO_1992}, we restrict our analysis here to such a slow particle creation regime, characterized by 
\begin{equation}
\Gamma_c \ll H.
\end{equation} 
Taking $\rho$ and $p$ as the energy density and pressure of the dominant matter component of the cosmic fluid, the curvature scalar in such a regime can be approximated by 
\begin{equation}
R \simeq \frac{1}{M_P^2}(1 - 3w)\rho. \label{g_field_2c}
\end{equation}

 \section{Cosmological baryogenesis in a radiation-dominated Universe}

\hspace{0.8cm}  We now consider the baryon asymmetry generation in the radiation dominated phase ($w\approx 1/3$) of the expanding Universe. 
As mentioned before, any baryon asymmetry generated above the GUT scale is washed out by cosmic inflation. Moreover, one expects a higher particle creation rate as the temperatures in the Universe get higher. Therefore, we assume an initial temperature just below the GUT scale
\begin{equation}
T_{\ast} \sim 10^{16} \ \mbox{GeV}.
\end{equation}
At very high temperatures, typical gauge groups and matter contents can yield a trace anomaly \cite{KLM_2003,STEINHARDT_ET_AL_2004}
\begin{equation}
(1 -3w_{\ast})\sim 10^{-1}-10^{-2}.
\end{equation}
Since $1 - 3w$ goes quickly to zero as the temperature drops, there is a decoupling temperature $T_0\ll T_{\ast}$, above the primordial nucleosynthesis temperature, for which $R_0 = 0$ ($w_0 = 1/3$). Recalling the definition $\D = n/s$, the baryon asymmetry at the decoupling temperature is, therefore, given by
 \begin{equation}
\D \simeq \frac{n_{B\ast}}{s_0}\left(\frac{\scalf_{\ast}}{\scalf_0} \right)^3 (1 + \coup R_{\ast}).
\label{baryon_number_densit2}
\end{equation}

We recall the expressions for the energy density for ultra-relativistic particles has the form (radiation) 
\begin{equation}
\rho(T) = \frac{\pi^2}{30} g_{\ast} T^4,
\end{equation}
where $g_{\ast}$ denotes the degrees of freedom of effective massless particles, and the entropy density
\begin{equation}
s(T) = \frac{2\pi^2 }{45}g_{s\ast}(T)T^3,
\label{entropy}
\end{equation}
where $g_{s\ast}(T)$ is the number of total degrees of freedom which contribute to the entropy of the Universe at a given temperature. For very high temperatures we can take $g_{s\ast}\sim g_{\ast}\sim 10^2$. From equations (\ref{g_field_2c}) and (\ref{baryon_number_densit2})-(\ref{entropy}) we finally obtain the baryon asymmetry at very high temperatures
 \begin{equation}
\D \simeq \frac{45}{60}\frac{n_{B\ast}}{T_{0}^3}\left(\frac{\scalf_{\ast}}{\scalf_{0}} \right)^3 \left[ \frac{30}{\pi^2g_{\ast}} + (1 - 3w_{\ast})\frac{\coup}{M_{P}^{2}} T_{\ast}^4 \right] ,
\label{baryon_asymmetry}
\end{equation}
while the highest value of the expected CP-violating phase $\theta'\equiv \coup R$ in (\ref{lagrangian_2}) is given by
\begin{equation}
\theta'(T_{\ast}) = \frac{\pi^2g_{\ast}}{30}(1 - 3w_{\ast})\frac{\coup}{M_{P}^{2}} T_{\ast}^4.
\end{equation} 

The primitive Universe is not expected to be exactly baryon symmetric, since there may be fluctuations in baryon and antibaryon distributions. Accordingly, in the absence of the mechanism ($\lambda R = 0$), one expects a baryon asymmetry from purely statistical fluctuations in baryon and antibaryon distributions to be of the order \cite{ALPHER_HERMAN_1953, ALPHER_HERMAN_1958} (see also\cite{RIOTTO_1998})
\begin{equation}
{\mathit{\Delta}}_{B}^{fluc} = \frac{45}{2\pi^2g_{\ast}}\frac{n_{B\ast}}{T_{0}^3}\left(\frac{\scalf_{\ast}}{\scalf_{0}} \right)^3\sim 10^{-39}.
\end{equation}
Thus, once the initial temperature and the trace anomaly are fixed, the only free parameter is the coupling constant $\lambda$. For the particular values of $T_{\ast}$ and $1-3w_{\ast}$ fixed above, a baryon asymmetry compatible with observation ($\D\sim 10^{-10}$) is obtained for a coupling constant with values in the range
\begin{equation}
1\ \mbox{GeV}^{-2} \lesssim \coup \lesssim 10^2\ \mbox{GeV}^{-2}.
\end{equation}
 The maximum value of the CP-violating phase that corresponds to these values of $\lambda$, on its turn, is of the order
\begin{equation}
\theta'(T_{\ast}) \sim 10^{29}.
\end{equation}

 \section{Cosmological baryogenesis in the post-inflationary reheating phase}

\hspace{0.8cm} We now consider the post-inflationary reheating phase due to the decay of a massive scalar field $\phi$ into relativist particles as it oscillates around the minimum of a quadratic potential. In this case, which corresponds to a matter-dominated epoch ($w = 0$), the curvature scalar reduces to
\begin{equation}
R =  \frac{1}{M_P^2} \rho,
 \label{g_field_2d}
\end{equation}
where, again, we are considering only the slow gravitationally induced particle production regime ($\Gamma_c\ll H$). Combining this expression with (\ref{baryon_number_densit}), we obtain
 \begin{equation}
\D \simeq {\mathit{\Delta}}_{B}^{fluc}\left( \frac{M_{P}^{2} + \coup\rho_{max}}{M_{P}^{2} + \coup\rho_{\ast}}\right) ,
\label{baryon_number_densit_b}
\end{equation}
where $\rho_{max}$ is the maximum energy density during the reheating phase, and $\rho_{\ast}$ is the reheating energy density.

The first case of interest is for $\rho_{max} \gg \rho_{\ast} \gg M_{P}^{2}/\coup$, which for $\coup \sim 1$-$10^{2}$ GeV can be attained for $\rho_{\ast}\sim 10^{35}$-$10^{39}\ \mbox{GeV}^4$. In this case, we have simply
\begin{equation}
\D \simeq {\mathit{\Delta}}_{B}^{fluc}\frac{\rho_{max} }{\rho_{\ast}} .
\label{baryon_number_densit_b1}
\end{equation}
Supposing again that ${\mathit{\Delta}}_{B}^{fluc} = n_B^{fluc}/s \sim 10^{-39}$, a baryon asymmetry compatible with observation and a minimum CP violation $\theta'(T_{\ast}) \sim 10^{29}$ are obtained for $\rho_{max} \sim 10^{64}$-$10^{68}\ \mbox{GeV}^4$.

Also for $\rho_{\ast} \lesssim M_{P}^{2}/\coup$ we obtain a similar result to that obtained in the previous section.

\section{Conclusions and outlook}

\hspace{0.8cm} In the present work, we have studied the implications of nonminimal curvature-matter couplings to the problem of the baryon asymmetry of the Universe. We have shown that for nonminimally coupled fermions (quarks) up to first order in the curvature, the field equations imply the nonconservation of the baryon current. Moreover, as suggested by QCD, an explicit CP-violating term is naturally expected in the Lagrangian of these theories. The fulfilment of the third of Skharov's conditions results from the associated gravitationally induced particle creation (in this case, quarks). This process amounts to the irreversible transfer of energy from the gravitational field to the created particles, and is thus an essentially out-of-equilibrium process. The nonconservation of the baryon current described here implies that the particle creation process associated with nonminimal curvature-matter couplings is also essentially baryon asymmetrical, in contrast with the standard general relativistic process. As a consequence, we conclude that all of Sakharov's conditions are already satisfied in the simplest extension of General Relativity, provided that the curvature scalar and its gradient do not vanish. 

In the present proposal, the particle creation pressure and the trace anomaly imply that a baryon asymmetry can be produced even in a radiation dominated Universe. In particular, the trace anomaly or the energy density of a massive scalar field can amplify the baryon asymmetry at very high temperatures, even for a very slow particle creation rate. As a consequence, a baryon asymmetry compatible with the observed value and a huge CP violation can be produced in a cosmological framework at a temperature just below the GUT scale ($\sim 10^{16}$ GeV) for a coupling constant with values in the range $1\ \mbox{GeV}^{-2} \lesssim \coup \lesssim 10^2\ \mbox{GeV}^{-2}$. A similar result is obtained in the post-inflationary reheating phase. 

Although we have restricted our analysis to a slow particle creation regime, it is clear that a fast particle creation could also produce interesting scenarios for baryogenesis. In that case, however, there is a risk of entropy overproduction and, as a consequence, a baryon asymmetry dilution. More importantly, such a fast particle creation process can be the source of a nonstandard cosmic dynamics (see, for instance, the discussion of the general relativistic case in \cite{LIMA_BASILAKOS_2012}). On the other hand, the mechanism proposed here could also be effective in other situations where very strong nonstationary gravitational fields might occur, such as in bubble collisions or domain wall collisions expected in a first-order cosmological QCD phase transition, which may happen in the case of a non-null chemical potential (\textit{e.g.} in the presence of sterile neutrinos). These possible scenarios will be addressed in future works.

\section*{Acknowledgements}

The authors thank the support from the Brazilian agencies CNPq, CAPES, Finep, and FAPERJ.

\end{document}